\documentclass[sigplan,10pt,pbalance]{acmart}
\renewcommand\footnotetextcopyrightpermission[1]{}
\usepackage{mathptmx}
\usepackage{graphicx}
\graphicspath{{./images/}}
\usepackage{algorithm}
\usepackage{xcolor,wrapfig}
\usepackage{listings,color}
\usepackage{booktabs}
\usepackage{siunitx}
\usepackage{url}
\usepackage{multirow}
\usepackage[noend]{algpseudocode}
\usepackage{xspace}
\usepackage{array}
\usepackage{pythonhighlight}
\usepackage{subcaption}
\usepackage{comment}
\usepackage{enumitem}
\usepackage{tikz}
\usepackage{import}
\usepackage[leftline=false,
            rightline=false,
            linewidth=.6pt,
            innerrightmargin=0pt,
            innerleftmargin=0pt,
            innertopmargin=0pt,
            innerbottommargin=0pt]{mdframed}
\usepackage{cleveref}

\usetikzlibrary{calc,arrows}

\makeatletter

\renewcommand\subsection{\@startsection{subsection}{2}{\z@}%
  {-.5\baselineskip \@plus -2\p@ \@minus -.2\p@}%
  {.25\baselineskip}%
  {\ACM@NRadjust\@subsecfont}}

\makeatother

\copyrightyear{2026}
\acmYear{2026}
\acmDOI{}
\acmISBN{}

\crefname{algorithm}{Algorithm}{Algorithms}
\crefname{listing}{Listing}{Listings}
\crefname{section}{\S}{\S\S}
\crefname{appendix}{Appendix}{Appendices}
\crefname{figure}{Figure}{Figures}
\crefname{table}{Table}{Tables}

\newcommand{\us}{\,$\mu$s\xspace}

\DeclareRobustCommand{\circled}[1]{\tikz[baseline=(char.base)]{
            \node[shape=circle,draw,inner sep=1pt] (char) {#1};}}

\lstdefinestyle{C}
{
    language=C++,
    basicstyle=\tiny\ttfamily,
    keywordstyle=\color{blue},
    stringstyle=\color{red},
    commentstyle=\color{black},
    morecomment=[l][\color{magenta}]{\#},
    numbers=left,
    numberstyle=\tiny\color{gray},
    stepnumber=1,
    numbersep=5pt,
    backgroundcolor=\color{white},
    showspaces=false,
    showstringspaces=false,
    showtabs=false,
    tabsize=2,
    captionpos=b,
    breaklines=true,
    breakatwhitespace=false,
    frame=single,
    escapeinside={@(*}{*)@}
}

\newcommand{\hOne}[1]{\section{#1}}

\newcommand{\hTwo}[1]{\subsection{#1}}
\newcommand{\hThree}[1]{\subsubsection{#1}}
\newcommand{\para}[1]{\noindentparagraph*{#1}}

\definecolor{dkgreen}{rgb}{0,0.75,0}
\newcommand{\result}[1]{{\color{dkgreen}#1}}
\renewcommand{\result}[1]{#1}

\newcommand{\work}{paper\xspace}

\begin{document}

\title[PonyTail: Profiling and Improving Service Tail Latency]{PonyTail: Profiling and Improving Service Tail Latency}

\author{Roee Wodislawski}
\affiliation{%
	\institution{Tel Aviv University}
	\city{Tel Aviv}
	\country{Israel}
}
\author{Adam Morrison}
\affiliation{%
	\institution{Tel Aviv University}
	\city{Tel Aviv}
	\country{Israel}
}

\begin{abstract}
Datacenter services must meet tight tail latency service-level objectives.
Research on improving tail latency focuses on reducing queuing delay by approximating optimal request scheduling policies.
As request scheduling nears optimality, the primary way to further improve tail latency is to reduce the service time of requests.
Unfortunately, datacenter services lack obvious hotspots for general service time optimization.

We point out a new optimization opportunity in targeting the \emph{tail service time} of services with dispersive service times, which are common in datacenters.
To show the feasibility and benefit of this approach, we present PonyTail, a performance analysis methodology and toolkit for analyzing service-level outlier behaviors associated with tail service time.
PonyTail helps performance engineers identify control-flow and microarchitectural patterns responsible for service time outliers, as well as their root causes.

We apply PonyTail to analyze five latency-critical services, including an in-memory database system and an online search engine.
Based on PonyTail's analysis, we optimize some of the services, improving their 99th percentile service time and latency by \result{10.7\%--46\%} and \result{21\%--$5\times$}, respectively.
\end{abstract}

\pagestyle{plain}
\maketitle

\newcommand{\xapian}{\texttt{xapian}\xspace}
\newcommand{\Xapian}{\texttt{Xapian}\xspace}
\newcommand{\silo}{\texttt{silo}\xspace}
\newcommand{\Silo}{\texttt{Silo}\xspace}
\newcommand{\moses}{\texttt{moses}\xspace}
\newcommand{\Moses}{\texttt{Moses}\xspace}
\newcommand{\shore}{\texttt{shore}\xspace}
\newcommand{\Shore}{\texttt{Shore}\xspace}
\newcommand{\sphinx}{\texttt{sphinx}\xspace}
\newcommand{\Sphinx}{\texttt{Sphinx}\xspace}
\newcommand{\masstree}{\texttt{masstree}\xspace}
\newcommand{\Masstree}{\texttt{Mastree}\xspace}
\newcommand{\imgdnn}{\texttt{img-dnn}\xspace}
\newcommand{\Imgdnn}{\texttt{Img-dnn}\xspace}

\newcommand{\StCo}{\textsc{StCo}\xspace}
\newcommand{\DyCo}{\textsc{DyCo}\xspace}
\newcommand{\uarch}{\textsc{uarch}\xspace}

\hOne{Introduction}\label{chap:introduction}

Datacenter services, from storage and data retrieval systems to machine learning applications, increasingly face tight tail latency service-level objectives (SLOs), such as the 99.9th percentile latency being a few milliseconds~\cite{TailAtScale,LowLatRedundancy,LatSLO50ms} or microseconds~\cite{KillerUsec,Shinjuku,Snap,Demikernel}.
These services form the backend of large-scale online platforms like social networks, web search, and e-commerce, which must respond to user requests quickly to improve user experience~\cite{TailAtScale} and revenue~\cite{LatencyRevenue1,LatencyRevenue2,LatencyRevenue3}.
Minimizing datacenter service tail latency is crucial, as each user request triggers hundreds of concurrent service calls~\cite{Dynamo,ScalingMemcache} and the slowest service response determines the request's overall latency~\cite{TailAtScale,DCComputer,tailbench-paper,DeathStarBench,NetflixMicroSvc}.

Tail latency is usually dominated not by the time it takes the service to process a request---its \emph{service time}---but by queuing delay in the operating system or service runtime system~\cite{Concord,ZygOS,Rubik}.
Queuing delay occurs when the request arrival rate momentarily exceeds the service's service rate, e.g., due to bursts of requests or service of a long-latency request blocking short-latency requests~\cite{Rubik,ZygOS,Persephone,Concord,Shinjuku}.

Most efforts to improve tail latency therefore target reducing queuing delay.
Queuing delay can be avoided by overprovisioning servers so individual server load is kept low, but this approach is wasteful and not cost-effective~\cite{EnergyPropForLatency,Quasar}.
Tail latency can be minimized in a resource-efficient manner by effective request scheduling (queue management).
Queuing theory has identified optimal work-conserving request scheduling policies%
\footnote{In a \emph{work-conserving} policy, servicing cores cannot be idle if there are pending requests.}
for minimizing tail latency~\cite{TailOptimalSched}.
Systems research has explored efficient mechanisms for implementing (or approximating) these policies---even for microse\-cond-scale service times---%
using work stealing~\cite{Shenango,ZygOS} and request preemption~\cite{Shinjuku,Concord,LibPreemptible}.
Compared to alternative (non-work-conserving) policies that leverage knowledge of the service time distribution~\cite{Persephone}, the near-optimal policies sustain higher throughput for the same tail latency SLOs.

With request scheduling subsystems tightly approximating optimal policies, however, the importance of improving service time grows.
The primary way to further reduce queuing delay under an optimal scheduling policy is to reduce request service time, as the time a request spends enqueued depends on the service times of all other pending requests~\cite{MorBook,TailOptimalSched}.

The problem is that service performance is \emph{already} heavily optimized, with hyperscale datacenter operators even performing automatic profile-guided optimization~\cite{Propeller,AutoFDO,BOLT,OCOLOS,Ripple} based on datacenter-wide profiling~\cite{ProfileWSC,AsmDB,GWP}.
Moreover, datacenter services typically do not have obvious hotspots that can be optimized to improve common-case service time~\cite{ProfileWSC}.
Driving tail latency improvements by improving service time will thus entail new optimization strategies.

\para{Opportunity: Tail optimization}
We point out a general optimization opportunity in targeting the \emph{tail service time} of services with dispersive (variable) service times, which are common in datacenters~\cite{GoogleClusterArch,MicroMicroSvc,tailbench-paper}.
In such services, it is meaningful to distinguish slow from average service times and to optimize the slowest requests---rather than optimizing hotspots that broadly impact a large fraction of requests.
Tail service time optimization is appealing for services with dispersive service times, as improving just the tail service times can have an outsized impact on queuing delay~(\cref{sec:motivation}).

Tail optimization targets \emph{service-level} factors, not external factors such as~(1)~hardware behaviors, e.g., resource competition from co-running programs~\cite{Quasar,BubbleFlux,Ubik,Heracles,Dirigent}, power limit-related throttling~\cite{Adrenaline,Rubik,EnergyPropForLatency,TimeTrader}, and energy management~\cite{PowerNap,DreamWeaver,KnightShift}; (2)~operating system sources like interrupt management~\cite{Reconciling,TalesOfTail,Shenango,ghOSt,Caladan}; or~(3)~runtime system activities such as garbage collection~\cite{TinyTail,TailLatSSD,Platinum}.
While variability in service time is often attributed to such external factors, many latency-critical services exhibit dispersive service times even when external performance fluctuations are controlled for~\cite{tailbench-paper}.
This indicates that their service time variability is due to service-level factors.

This \work shows the feasibility and benefit of tail service time optimization. We make two main contributions.

\para{\circled{1} PonyTail (\cref{chap:framework})}
We present PonyTail, a performance analysis methodology and toolkit for developers to explore tail service time optimization possibilities.
PonyTail's goal is to assist performance engineers in analyzing service-level outlier behaviors responsible for tail service time in cases where hotspot-oriented profiling is less applicable.
Because the outlier behaviors behind tail service times occur infrequently (e.g., 1-in-1000 for the 99.9th percentile service time), a traditional profiler would either miss them or rank them as insignificant.
Example PonyTail use cases include analysis of service time variability in a specific request type and identification of slow requests in a mix of request types.%
\footnote{In contrast, if tail service time is a priori known to result from a specific slow request type (such as range scans vs. point queries in a key/value store~\cite{Persephone}), it can usually be analyzed by profiling a workload with only slow requests.}

PonyTail's methodology seeks to attribute outlier service times to the following service-level causes, which together characterize all possible reasons for a tail request being slower than typical requests.
The relevant requests might have a unique \emph{architectural (code) footprint}---i.e., they take a distinct execution path, either statically (different code) or dynamically (e.g., executing more loop iterations than in average requests).
Alternatively, outlier requests might have a distinct \emph{microarchitectural footprint}, such as a certain load instruction incurring a cache miss instead of a hit.

PonyTail's toolkit enables developers to identify outlier footprints and their root causes.
It provides a familiar profiler workflow, consisting of online recording of every request's footprints, followed by offline analysis that identifies unique aspects of slow request footprints.
To help subsequent analysis of these unique aspects' root causes, PonyTail supports request-level collection of user-specified performance counter measurements with a similar workflow.

PonyTail's data collection leverages modern capabilities of Intel processors, such as Processor Trace~\cite{sdm-manual} and top-down microarchitecture analysis performance counters~\cite{sdm-manual}.
It also requires minimal service source code modification to demarcate request boundaries by invoking PonyTail methods.

Our evaluation~(\cref{chap:eval}) shows that PonyTail's data collection has moderate overhead---increasing mean service time by \result{9\%--24\%}, depending on the service time scale---but does not change the shape of the service time distribution.
While this overhead precludes datacenter-wide, 24/7 deployment of PonyTail data collection, we believe that it is acceptable for intermittent or manual data collection for performance analysis.
PonyTail data collection can also be run on multiple server machines (i.e., service instances), thereby minimizing disruption to each individual instance while still collecting sufficient data to capture tail events.

\para{\circled{2} Case studies (\cref{chap:cases})}
To demonstrate the merit of the PonyTail methodology and toolkit, we apply them to analyze the services from the TailBench suite~\cite{tailbench-paper}, a collection of latency-critical services spanning a wide range of application domains and tail latency SLOs.
Seven out of the eight services are C++ programs, of which we analyze the five services that exhibit dispersive service times.

PonyTail successfully identifies that service time outliers in these services are correlated with distinct microarchitectural and code footprints (both static and dynamic).
Based on insights obtained from examining these footprints and their root causes, we optimize some of the services, each in a different way---from low-level optimization to configuration tuning.

Our tail optimizations improve the 99th percentile service time of TPC-C transactions in the in-memory database system \silo~\cite{silo-paper} by \result{10\%} and its maximum service time by \result{$4.3\times$}.
Impact on 99th percentile latency depends on workload queuing delay: it improves by \result{21\%} at low-to-moderate loads (where the baseline 99th percentile latency is at most $12\times$ the mean service time) and by up to $3\times$ at higher loads.
For the real-time translation service \texttt{moses}~\cite{moses-paper}, we find that tail service time is mainly due to misses in a service-internal cache.
Improving the cache's hit rate (e.g., via warmup) results in reductions of $53\times$ and $22\times$ in the 99th percentile service time and latency, respectively.

In the online search engine \xapian~\cite{xapian-website} and the speech recognition service \sphinx~\cite{sphinx-paper-2}, we find that tail service time is due to factors such as input-dependent loops which dominate service time, and thus cause service time outliers for certain inputs.
In these cases, PonyTail's unique tail-oriented analyses lead to discovery of optimizations that improve service times across the board.
Some of these optimizations may be independently interesting, as they target under-explored microarchitectural mechanisms.
Our optimizations improve \xapian and \sphinx service times by \result{46\%} and \result{10.7\%}, respectively, yielding improvements of up to $5\times$ and $2\times$ in latency~(across the board).

Overall, this \work shows that opportunities to further reduce tail latency exist and that tail service time optimization---facilitated by PonyTail---can unlock this potential, thereby complementing existing optimization approaches.

\hOne{Background}\label{chap:background}

This section provides background on performance analysis on Intel CPUs.
Familiar readers can skip this section.

\para{CPU structure}
Like all modern high-performance CPUs, Intel CPUs employ a superscalar out-of-order (OoO) microarchitectural design~\cite{SuperScalarUarch}.
``Superscalar'' means that the processor can initiate execution of multiple instructions (from a single instruction stream) per clock cycle~\cite{hennessy2011computer,johnson1991superscalar}.
``Out-of-order'' (or \emph{dynamic}) means that the processor determines instruction execution order according to the dynamic availability of instruction inputs and hardware execution units~\cite{tomasulo1967efficient}, executing instructions possibly out of program order while retaining the appearance of a sequential execution.
A processor core consists of a \emph{frontend}, which fetches instructions from memory, and a \emph{backend}, which executes instructions.
The frontend \emph{decodes} fetched instructions into internal micro-operations ($\mu$-ops), which are subsequently \emph{dispatched} to the backend, where they await execution.
The backend is responsible for executing a fixed-size sliding instruction window of the latest fetched $\mu$-ops.
$\mu$-ops \emph{retire} in program order, exiting the instruction window once they have executed and committing their results to architectural state.
The backend computes dependencies between the $\mu$-ops and schedules their execution (by \emph{issuing} them to an execution unit) regardless of their position in the instruction window, based only on availability of hardware execution resources and dependency constraints.

The CPU relies on branch prediction~\cite{hennessy2011computer} to avoid stalling instruction fetching when encountering a branch instruction.
Once the branch executes, the prediction's correctness is validated.
If the prediction was correct, the branch instruction can eventually retire.
Otherwise, on a misprediction, the backend's state is rolled back to its last known valid state (which was checkpointed upon making the prediction), thereby squashing the mispredicted branch and all subsequent (mispredicted) $\mu$-ops in the instruction window

\para{Top-down microarchitecture analysis~\cite{TMA-paper}}
TMA is a methodology for identifying hardware-related bottlenecks.
It uses performance monitoring events to measure utilization of \emph{pipeline slots}, which represent the CPU resources required to process a $\mu$-op.
Each pipeline slot available during the program's execution is classified into one of four categories.
If the slot is empty (stalled), it is classified according to the stall reason:~(1)~frontend (unable to fill the slot, e.g., due to an instruction cache miss) or~(2)~backend (unable to process the slot, e.g., due to an unavailable resource).
If the slot is non-empty (utilized), it is classified according to whether it~(3)~was mispredicted or~(4)~eventually retired.
TMA can further break down each category into finer-grained metrics.

The following performance monitoring mechanisms are provided by the performance monitoring unit (PMU) in Intel CPUs.
They can be enabled by a profiler without cooperation from the analyzed program.

\para{Processor event-based sampling~\cite[\S~20.9]{sdm-manual}}
PEBS enables efficient, precise sampling of hardware events.
In contrast to traditional sampling, recording the execution context of each sample (e.g., timestamp and program counter) is done by the CPU.
This information is recorded in an in-memory buffer, with an interrupt raised only when the buffer fills up.

\para{Processor Trace~\cite[\S~33]{sdm-manual}}
PT captures a control-flow trace of the program into an in-memory buffer in compact format.
For every executed branch (except unconditionally taken branches), it records whether the branch was taken or not, its target (for an indirect branch), and (optionally) a timestamp.

\hOne{Motivation} \label{sec:motivation}

Service tail latency SLOs are specified with respect to \emph{response time}, which is the sum of a request's queuing delay and service time.
Queuing delay dominates response time unless the system operates well below capacity, and is therefore the main target for optimization.
As a result, implementations of tail-optimal request scheduling policies have seen continuous efficiency improvements, to the point that further tail latency reductions will require improving service times---but at the same time, it is becoming harder to find software optimizations that broadly improve service times~(\cref{sec:motivate-need}).
To address these trends, we argue for optimizing tail service time, as it can have an outsized impact on response time~(\cref{sec:motivate-tailsvctime}).

\hTwo{Need for tail service time optimization} \label{sec:motivate-need}

Queuing theory has established an optimal configuration for minimizing tail latency: a single request queue managed with a tail-optimal scheduling policy, which depends on the nature of the service time distribution~\cite{TailOptimalSched}.
For low-to-moderately dispersive service times, the optimal policy is first-come-first-served (FCFS)~\cite{FCFS-optimal}.
For highly dispersive distributions, an optimal policy is processor sharing (PS)~\cite{PS-optimal}, where each request receives an equal fraction of CPU capacity.

Implementations of these optimal configurations must address problems that are abstracted away in the queuing theory model, such as multi-core synchronization of a global queue and how to time-share cores between requests in PS.
Over time, implementations have steadily addressed the overheads caused by these problems~\cite{ZygOS,Shenango,Persephone,Shinjuku,Concord,Caladan,LibPreemptible}.
State-of-the-art request schedulers can now efficiently implement (or closely approximate) the optimal configurations even for microsecond-scale service times.
For example, the overhead of the Concord~\cite{Concord} preemptive scheduler (which approximates PS) on 500\us requests is only 1.5\%--5\% (depending on the hardware) even for a scheduling quantum of 2.5\us.
Overall, request scheduling overhead is now negligible, except perhaps for single- or double-digit \us requests.
And even there, scheduling overhead is continuously decreasing.

Given an optimal request scheduling policy implemented with minimal overhead, \emph{the primary way to further improve tail latency is by improving service times}, as queuing delay depends on the service times of pending requests.
In FCFS, a request's queuing delay is the sum of the service times of the requests in front of it in the queue.
Clearly, reducing the pending requests' service times will reduce the request's queuing delay.
In PS, a request's response time depends on the fraction of CPU capacity it receives, which depends on the number of pending requests.
If some pending requests finish faster, the request's CPU share will increase and it will complete sooner.

While it would be ideal to improve service times across the board by speeding up a large fraction of requests, it is becoming harder and harder to find software optimizations with such an effect.
Services often have no obvious hotspots to optimize, to the point that optimizing individual services is considered not cost-effective~\cite{ProfileWSC}---which has driven datacenter operators to perform automatic profile-guided optimization~\cite{Propeller,AutoFDO,BOLT,OCOLOS,Ripple}.
Recent research on broad optimization of datacenter services has turned to proposing hardware-based optimizations~\cite{ProfileWSC,Thermometer,Whisper,Twig}, which cannot be used today.
We conclude that \emph{a new software optimization strategy is needed: tail service time optimization---i.e., optimizing the slowest requests instead of most or all requests.}

\hTwo{Potential of tail service time optimization} \label{sec:motivate-tailsvctime}

\begin{figure*}
    \centering
        \includegraphics[width=0.5\columnwidth]{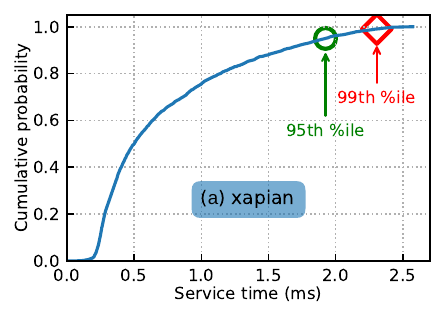}
        \includegraphics[width=0.5\columnwidth]{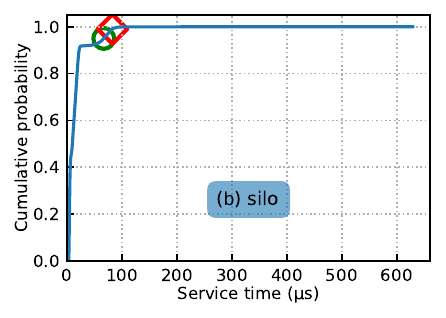}
        \includegraphics[width=0.5\columnwidth]{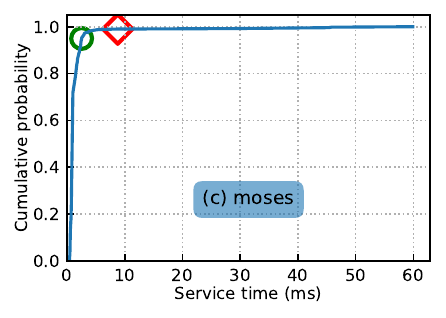}
        \includegraphics[width=0.5\columnwidth]{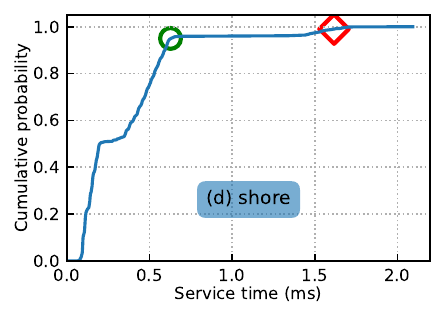}\\
        \includegraphics[width=0.5\columnwidth]{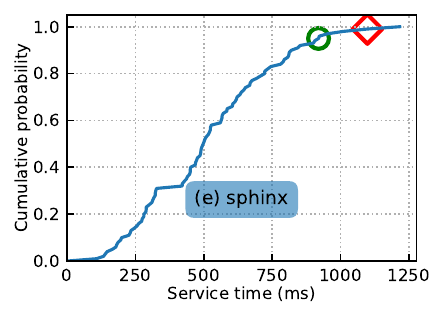}
        \includegraphics[width=0.5\columnwidth]{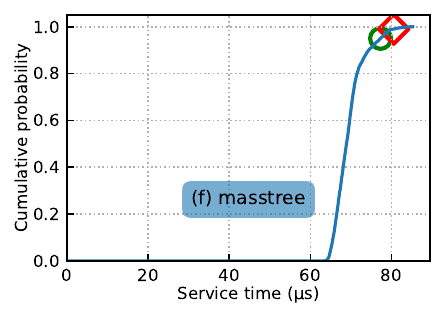}
        \includegraphics[width=0.5\columnwidth]{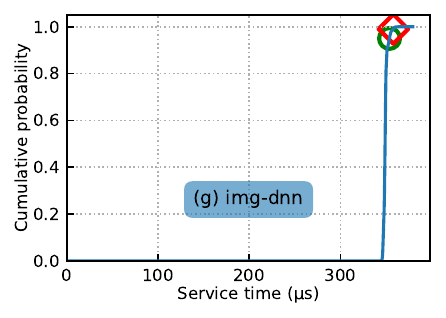}
        \caption{Service time cumulative distribution function (CDF). Green circles and red diamonds mark 95th and 99th percentiles. The CDF of \sphinx is obtained over a longer measurement period than used in its analysis~(\cref{subsec:sphinx}): 100 requests vs. 20 requests. The \masstree and \imgdnn services are not analyzed in the paper due to their low-dispersion service times. We describe them briefly for completeness: \masstree is an in-memory key-value store and \imgdnn is a C++ application that utilizes the OpenCV library to perform handwriting recognition.}
        \label{fig:xapian-time} \label{fig:silo-time} \label{fig:moses-time} \label{fig:shore-time} \label{fig:sphinx-time} \label{fig:masstree-time} \label{fig:img-dnn-time}
\end{figure*}

To show the potential of optimizing tail service time, we simulate the impact of hypothetical tail service time optimizations on the 99.9th percentile response time of a latency-critical service running on an ideal server.
We consider the (dispersive) service time distributions of four latency-critical services: \xapian, \silo, \moses, and \shore (described in~\cref{chap:cases}, \cref{fig:silo-time}) whose service times range across different scales, from tens of microseconds to tens of milliseconds.

We simulate a typical server with 16 worker cores using the optimal configuration of a single queue managed with a tail-optimal scheduling policy.
We determine the optimal policy for each service by simulating both FCFS and PS, as which of them is optimal depends on the level of distribution dispersion~\cite{ZygOS}.
We simulate the system until achieving 95\% confidence intervals of 5\% error in the obtained results.

We compare each service to hypothetical tail-optimized versions of itself, denoted $\text{opt}\left(\geq p,-d\%\right)$, in which the requests responsible for the $p$-th percentile (or worse) service times have been sped up by $d$\%.
We simulate $\text{opt}\left(\geq p,-d\%\right)$ by sampling an arriving request's service time from the service's baseline distribution and reducing it by $d$\% if it exceeds the baseline $p$-th percentile.

\begin{figure}[t]
    \centering
    \includegraphics[width=\columnwidth]{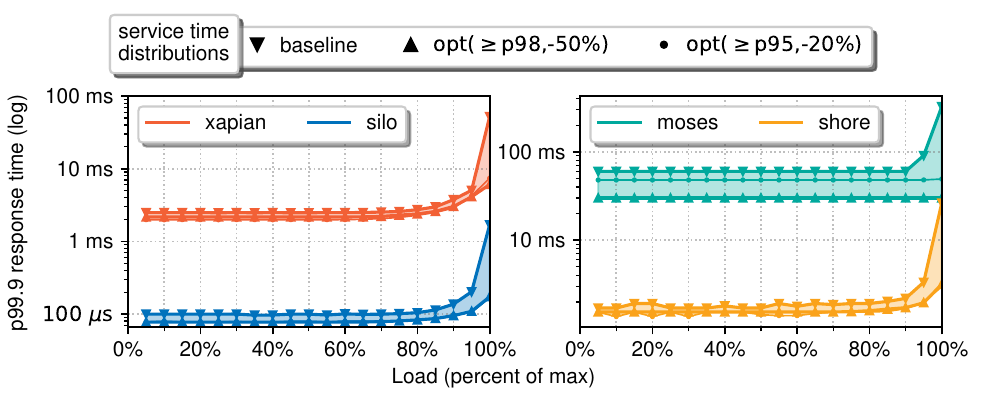}
    \includegraphics[width=\columnwidth]{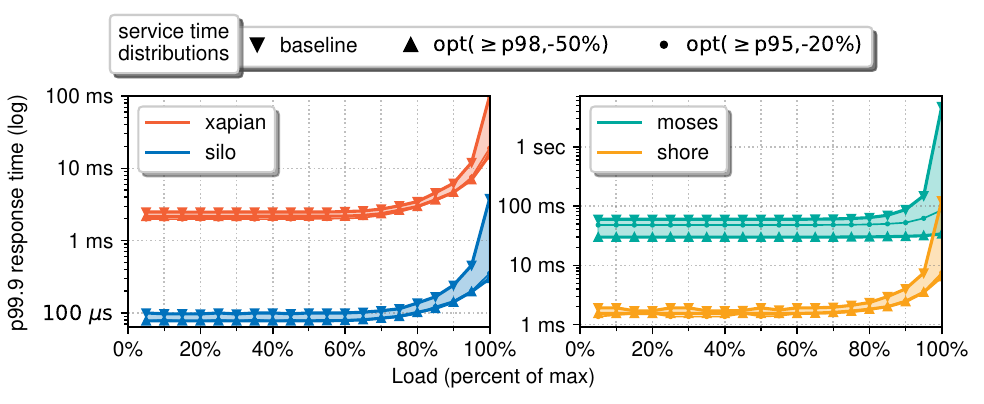}
\caption{\textbf{Tail service time impact on tail response time (including queuing delay) for an optimal FCFS scheduling policy (top) and a processor sharing scheduling policy (bottom).} The figure shows how a hypothetical reduction of 50\% or 20\% of the 98th or 95th percentile service times, respectively, impacts the 99.9th percentile response time of four latency-critical services (with different service time scales) at various loads.} \label{fig:Fifo_lats_vs_load}
\end{figure}

\Cref{fig:Fifo_lats_vs_load} compares each service to its $\text{opt}\left(\geq p98,-50\%\right)$ and $\text{opt}\left(\geq p95,-20\%\right)$ versions, which model optimization levels similar to what our case studies~(\cref{chap:cases}) achieve.
We report results for FCFS (which is optimal for every service) and PS. The results for both are qualitatively similar.
The figure compares the 99.9th percentile response time for Poisson request arrivals with the same mean request rate, $\lambda$, shown as a fraction of the baseline's service rate of $\mu \cdot w$, where $\mu$ is the baseline's single-worker mean service time and $w=16$ is the number of workers.
(A Poisson arrival process models the bursty, independent nature of datacenter traffic~\cite{QueuingFirstApproach}.)

There are several takeaways from~\cref{fig:Fifo_lats_vs_load}:
\circled{1}~\emph{Tail service time optimization can yield disproportionate improvements in tail response time and throughput.}
Although the mean response time of \xapian, \silo, and \shore improves by $\leq 6.5$\% due to the tail service time optimizations, their 99.9th percentile response time at 95\% load improves by 20\%, 47\%, and 42\%, respectively.
Likewise, the optimized \silo and \shore (respectively, \xapian) can handle 10 (respectively, 5) percentage points higher request rates with the same tail latency SLO as their baselines at 80\% load.
(We discuss \moses separately below.)

\circled{2}~\emph{Moderate speedups can be effective.}
For \xapian, \silo, and \shore, the impact of speeding up the 5\% slowest requests by 20\% is comparable to that of speeding up the 2\% slowest requests by an aggressive 50\%.
This is because optimizing the 98th percentile by 50\% ``shifts'' lower percentiles upwards, making them the new tail.

\circled{3}~\emph{Speeding up the long tail can be crucial}.
The tail of \moses' service time distribution is much heavier than that of the other services, with $30\times$--$85\times$ greater outlier service times.
Consequently, \moses benefits significantly more from $\text{opt}\left(\geq p98,-50\%\right)$ than from $\text{opt}\left(\geq p95,-20\%\right)$.
Despite improving mean response time by 16\% and 7.5\%, respectively, these hypothetical versions improve tail response time at low load by 50\% and 20\%, and at 95\% load, by 70\% and 40\%.
Both versions can thus sustain their baseline's maximal load with a better tail latency SLO.
This scenario motivates optimizing the far tail of the service time distribution.

\hOne{PonyTail}\label{chap:framework}

PonyTail is a performance analysis methodology~(\cref{sec:method}) and toolkit~(\crefrange{sec:design}{sec:impl}) for assisting performance engineers in analyzing service-level outlier behaviors responsible for tail service time and exploring tail optimization possibilities.
Hotspot-oriented profiling is less applicable for these tasks, because the relevant behaviors may occur infrequently (e.g., 1-in-1000 for the 99.9th percentile service time) and so get missed or ranked as insignificant.

\hTwo{Methodology and workflow}\label{sec:method}

PonyTail's methodology seeks to attribute outlier service times to one or more of the following three cases, which together characterize all possible reasons for a tail request being slower than typical requests.

\para{\circled{1} Different static code (\StCo)}
In addition to typical code paths, a tail request might exercise other code paths whose execution slows request processing.
Examples of this scenario include invocation of garbage collection code or distinct request types, some of which are slow.

\para{\circled{2} Same static code, different dynamic behavior (\DyCo)}
Tail requests might execute exactly the same static code as average requests, but in a different dynamic way which slows down the request.
An example of this case is a loop whose iteration count depends on a request parameter $x$, which results in high service times for requests with high values of $x$.

\para{\circled{3} Same static and dynamic architectural behavior, different microarchitectural behavior (\uarch)}
Tail requests might be slowed down by microarchitectural factors.
For example, certain memory load instructions might hit in the last-level cache (LLC) in average requests but miss in tail requests.

\para{Workflow}
PonyTail's toolkit is designed to facilitate identifying which of the above cases affect the service and determining their root cause using the manual workflow described in~\crefrange{sec:workflow1}{sec:workflow3}.

\hThree{Characterization} \label{sec:workflow1}

To determine which of \StCo, \DyCo, and \uarch are present, we first characterize service times in terms of the \emph{code} and \emph{microarchitectural footprints} of their respective requests.
A code footprint specifies which code locations (basic blocks) a request executes and how frequently.
A microarchitectural footprint is a request-level TMA breakdown of pipeline slot usage.

Comparing request footprints allows classifying requests into homogeneous \emph{groups} of requests with similar footprints, i.e., similar code flow or microarchitectural bottlenecks.
This classification reveals which of \StCo, \DyCo, and \uarch occur at each service time percentile range:
\StCo occurs if the range is associated with a code footprint-defined group; otherwise, \DyCo occurs.
\uarch occurs if the range is associated with a microarchitectural footprint-defined group.

PonyTail assists characterization by providing a graphical representation of code and microarchitectural footprints across service times, which enables visually identifying request groups~(\circled{1} in ~\cref{sec:design:report}).
PonyTail describes code footprints in terms of \emph{dynamic basic blocks} (DBBLs).
A DBBL is a sequence of consecutive retired instructions $\tau=i_1,\dots,i_k$ in a program execution $E$ such that no branch in $E$ jumps to any instruction in $\tau$ but $i_1$  and no \emph{taken} conditional branches occur in $\tau$ except for $i_k$.%
\footnote{Instructions other than $i_k$ can be branches, but in $E$, their execution always falls through (branch not taken).}
These conditions are the dynamic analogues of the ``no branches in'' and ``no branches out'' definition of a static basic block.
PonyTail uses DBBLs because their boundaries can be efficiently extracted from Processor Trace data, whereas basic blocks require disassembling the service binaries~(\cref{sec:impl:report}).

\begin{figure}
        \centering
        \includegraphics[width=0.5\columnwidth]{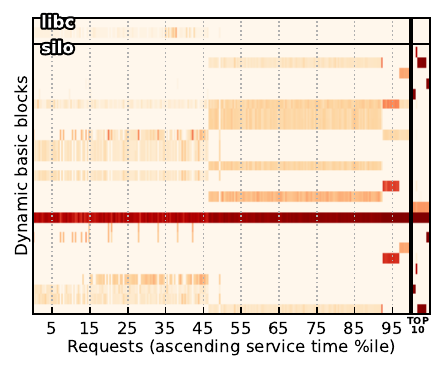}%
        \includegraphics[width=0.5\columnwidth]{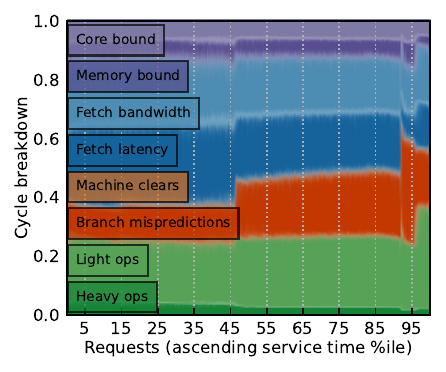}\\
        \begin{minipage}[t]{0.5\columnwidth}
        \centering
        \subcaption{$\left<35\%,10,10\right>$-DBBL heatmap.} \label{fig:silo-char-hmap}
        \end{minipage}%
        \begin{minipage}[t]{0.5\columnwidth}
        \centering
        \subcaption{TMA metrics.} \label{fig:silo-char-tma}
        \end{minipage}%
        \caption{\textbf{PonyTail analysis of \silo requests.} Distinct request groups are revealed (e.g., percentiles $<$45, 45--90, 90--95).}
    \label{fig:silo-char}
\end{figure}

PonyTail visualizes code footprints by plotting a heatmap of request-level (ordered by service time) DBBL \emph{hotness}, which is the fraction of time the request spends in this DBBL and any function it calls, with DBBLs identified by their location in the program's address space
(with black horizontal lines separating different modules).
The user can plot only DBBLs with hotness of at least $h$\% and that execute at least $n$ times in each request. We refer to such a plot as a $\left<h,n\right>$-DBBL heatmap, or a $\left<h,n,k\right>$-DBBL heatmap if it also zooms in on the slowest $k$ requests.
\Cref{fig:silo-char-hmap} shows an example with several groups (e.g., below and above the 45th percentile).

PonyTail visualizes microarchitectural footprints by plotting request-level TMA breakdowns, ordered by service time.
\Cref{fig:silo-char-tma} shows an example, again with noticeable groups.

\hThree{Exploration} \label{sec:workflow2}
This step aims to pinpoint the root causes of tail service time.
If the relevant service time percentile range is uniquely associated with a request group identified above, we need to identify the reasons for software (\StCo) or hardware (\uarch ) inefficiency.
PonyTail assists this process by enabling standard performance analysis---such as sampling hardware events and identifying hot code locations---of only that group~(\circled{5} in~\cref{sec:design:report}).
Limiting to a specific group sidesteps the problem a standard profiler would have.
Its problem is not that events in tail requests are not sampled---they are, albeit with lower probability.
But because these samples are infrequent, they are ranked as insignificant.
PonyTail's zooming in only on relevant samples thus creates a request-level profile analysis.

A different problem must be solved if tail service times are not defined by a group (\DyCo).
Here, we are not looking for bottlenecks but rather dynamic behaviors (typically loops) that cause slow service time.
To this end, beyond PonyTail's group-level performance analysis, PonyTail can also identify \emph{key} DBBLs whose execution time or count is correlated with service time and their call stacks~(\circled{2} in~\cref{sec:design:report}),
as well as generate plots depicting the correlation across all
requests~(see~\cref{fig:xapian-bbl} for an example).

Exploring such DBBLs is typically prioritized by their call stack depth, to allow for root cause detection in higher code levels.
PonyTail also helps to explore loop behavior by providing request-level loop iteration counts~(\circled{4} in~\cref{sec:design:report}).

\hThree{Optimization} \label{sec:workflow3}
Insights from the analysis hopefully lead to optimizations that improve tail latency.
Importantly, optimizations made in the \DyCo case might impact the entire request group (but not necessarily).
If the optimization addresses a loop-related hardware bottleneck, for example, then every request in the group benefits.
However, if the optimization parallelizes loop iterations only for certain request arguments, only that part of the group benefits.

\hTwo{Design}\label{sec:design}

\begin{figure}
    \centering
    \includegraphics[trim=0.5cm 17cm 0.5cm 3cm, width=\columnwidth]{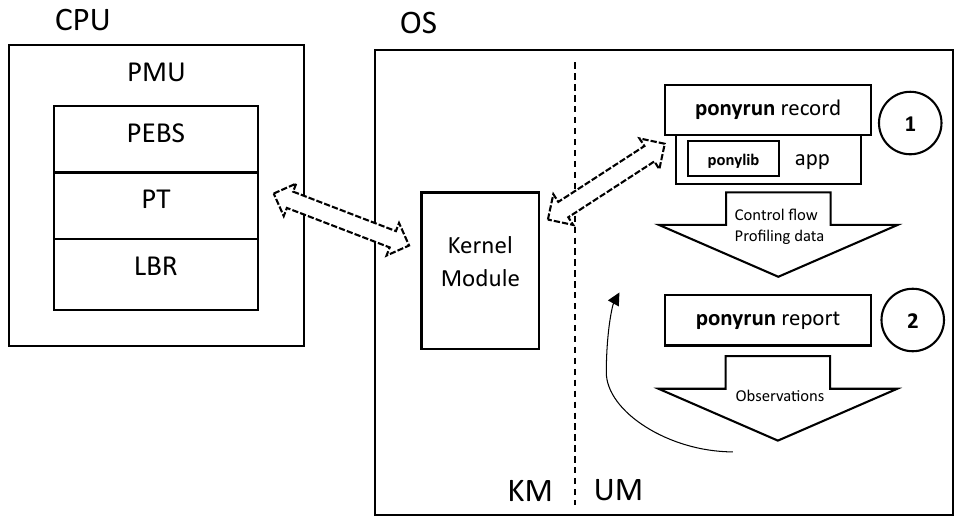}
     \caption{\textbf{PonyTail design.}}
     \label{fig:design}
\end{figure}

PonyTail's toolkit provides a familiar profiler interface, consisting of online recording of performance monitoring data~(\cref{sec:design:record}) followed by offline analysis of the data to help the user identify unique aspects of slow requests~(\cref{sec:design:report}).
\Cref{fig:design} depicts the toolkit's design. It consists of a profiler-like tool, \texttt{ponyrun}, which can run in ``record'' mode for data collection or in ``report'' mode to analyze collected data.
PonyTail data collection relies on request boundaries being dynamically demarcated by the service via calls into \texttt{ponylib}, a dynamically linked library.

\hThree{Data collection (\texttt{record} mode)} \label{sec:design:record}
Most of the data PonyTail collects is identical to that collected by existing profilers, such as the Linux \texttt{perf} tool~\cite{PerfTool} and Intel's VTune~\cite{VTune}.
Data collection has three standard modes: (1)~\emph{tracing}, which collects a control-flow trace using Processor Trace (PT); (2)~\emph{sampling}, which samples user-specified hardware events using processor event-based sampling (PEBS); and~(3)~\emph{statistics}, which collects request-level performance counter data for user-specified counters, which can be useful during the exploration step~(\cref{sec:workflow2}).

\begingroup
\setlength{\columnsep}{12pt}

PonyTail's main unique aspect is its analysis, which breaks the program-level data into request-level data, even in sampling mode.
To enable this, PonyTail requires the analyzed service to dynamically demarcate (with \texttt{pony\-lib} calls)
the periods in which a request executes, i.e., when it starts/completes and stops/resumes due to service-level scheduling, if any~(\cref{tab:ponylib}).%
\footnote{Process preemption is handled automatically by the kernel performance monitoring interface.}

\endgroup

\begin{table}
    \centering
    \small
    \begin{tabular}{@{}l|l@{}}
    \toprule
    \textbf{Operation} & \textbf{Invoked when?} \\
    \midrule
    \texttt{req\_begin()} & Start of new request.\\
                          & Returns a unique req. id. \\
    \texttt{req\_stop(id)} & Request \texttt{id} is preempted. \\
    \texttt{req\_cont(id)} & Request \texttt{id} resumes. \\
    \texttt{req\_done(id)} & Request \texttt{id} completes. \\
    \bottomrule
    \end{tabular}
    \caption{\textbf{Ponylib interface.}}
    \label{tab:ponylib}
\end{table}

\texttt{Ponylib} serves several purposes.
First, it collects request-level counts of performance counters by accumulating their difference before and after request activity periods.
To collect TMA metrics, \texttt{ponylib} tracks special-purpose counters with TMA metrics available in recent Intel processors, which enable constructing a full TMA breakdown during analysis.
(With the release of the Ice Lake microarchitecture, Intel introduced built-in support for highest-level metrics of the TMA method level. With Sapphire Rapids, the built-in support was extended to include second-level metrics~\cite{sdm-manual}.
Until these additions, it was required to interpolate readings of different performance counters to estimate each metric. As the number of performance counters in use is limited, it was not possible to achieve an overall view of these levels in the same execution.)
It also implements the request-level tracking of user-specified performance counters option.
Finally, \texttt{ponylib} records each request's activity time intervals.
This data enables PonyTail's analysis to associate samples with requests~(\cref{sec:impl:record}).

\hThree{Analysis (\texttt{report} mode)} \label{sec:design:report}

PonyTail analysis takes collected data as input and produces one of the following user-specified reports.
Crucially, reports can be limited to consider only a certain request group (instead of all requests).
Groups are defined by specifying a DBBL, which means ``consider only requests in which this DBBL executes.''
A DBBL is specified as a tuple of a dynamic shared program object (DSO) name and an offset inside it.

Unless noted otherwise, the analyses process tracing data.

\para{\circled{1} Characterization}
This report produces the visualizations of request code and microarchitectural footprints that are used in our workflow's initial characterization step~(\cref{sec:workflow1}).

\para{\circled{2} DBBL-Info}
This report helps to identify DBBLs  whose execution time or frequency is correlated with service time.
It provides a list of all DBBLs in the desired request group whose hotness exceeds a user-specified threshold.
Output information includes hotness, average observed call stack depth, and execution count and time, for which PonyTail also displays the squared Pearson correlation coefficient between the DBBL and request service time.
See~\cref{table:silo-gc-bbl-info} for an example.

\para{\circled{3} DBBL-Follow}
When a DBBL's execution count or time is suspected to cause high service time in the request group, this report can be used to plot the correlation across all requests.

\para{\circled{4} Loop-Follow}
To understand the behavior of loops that cause high service time, this report displays the frequency with which a given loop performs $i$ iterations.
The followed loop is specified with a DSO and a list of tuples $\left<o, t_o, l_o\right>$, where $o$ is an offset in the DSO, $t_o$ indicates whether $o$ is the address of a taken branch or its target, and $l_o$ indicates whether this jump indicates loop start, iteration, or termination.

The analysis tracks the recorded control-flow.
Every time the ``start'' taken branch is encountered, it starts counting ``iteration'' taken branches until the ``termination'' branch (if specified) or an unspecified branch is encountered.
Note that an inherent limitation of the described implementation is that loops whose body include function calls or other branches are not supported.

\para{\circled{5} Profile}
This analysis summarizes data recorded in sampling or statistics mode in the form of sampled event counts or per-request counts, respectively.
It considers only data relevant to the specified request group (if any), i.e., samples taken during group requests or statistics collected for them.

\hTwo{Implementation}\label{sec:impl}

We implement \texttt{ponyrun} by extending Linux's \texttt{perf} tool.
For data collection, \texttt{perf} can essentially be used as-is.
Recording with \texttt{perf} saves engineering effort, as it takes care of managing the PMU during context switches, decoding recorded events and augmenting them with timing information, mapping addresses to symbols, etc.
For efficient analysis, we use the \texttt{dlfilter} interface of \texttt{perf}, which allows processing the recorded data through a custom filter provided by a shared object file.
PonyTail's analyses are implemented as such a filter in C++ in 1450 lines of code (LOC).
The rest of \texttt{ponyrun} is implemented in Python (2090\,LOC).
\texttt{Ponylib} is implemented in C++ (670\,LOC), using the \texttt{jevents} library~\cite{jevents-github}.

\hThree{Data collection} \label{sec:impl:record}

\para{Control-flow tracing}
PonyTail uses Intel PT (supported by \texttt{perf record}) to capture the service control-flow trace.
We configure Intel PT to record all taken branches (both conditional and indirect branches) and to additionally log timing information, but at a reduced rate, so the overall recording rate is tens of MBs/second.
To attribute control-flow events to specific requests, \texttt{ponylib} injects entries demarcating request activity into the PT log by issuing \texttt{PTWRITE} instructions.

\para{Event-based sampling}
PonyTail uses PEBS via \texttt{perf record} to sample CPU context and call stack for every predetermined number of occurrences of the specified hardware events.

\hThree{Analysis} \label{sec:impl:report}

The analysis begins with \emph{DBBL extraction}.
This task is complicated by the fact that \texttt{perf} configures PT not to log non-taken branches, to curb the PT data volume.
PonyTail uses the sources and targets of taken branches to ``fill in the gaps''---i.e., to distinguish between a dynamic instruction sequence forming one DBBL and a DBBL that falls through to another.
\Cref{sec:appendix} shows the algorithm.

Next, we describe key implementation aspects.

\para{DBBL-Info}
PonyTail determines the hotness of DBBL $\tau$ for request $r$ based on periodic samples of stack traces $S_r$ captured during $r$'s execution, as follows:
$$H_{\tau,r} = \dfrac{\left|\left \{ s \in S_r : \exists a \in s.stack\; \text{s.t.}\; a \in \tau \right\} \right|}{\left| S_r \right|}$$

The analysis synthesizes the stack samples from the PT control-flow trace, as the trace contains the data required to obtain the call stack at any point in time.

\para{Profile}
To enable the analysis to associate samples with requests, we match the monotonic timestamp  and thread ID \texttt{perf} provides for each sampled PEBS event with the information recorded by \texttt{ponylib} about request activity.

\hTwo{Limitations}

PonyTail assists manual optimization instead of automatically addressing performance issues~\cite{DMon,AutoFDO,APT-GET,BOLT,OCOLOS}.
This is a conscious design choice, whose rationale is discussed in~\cref{sec:literature}.

Due to the complex environment of datacenter services, it is difficult to profile their real-world behavior outside of the production deployment~\cite{DMon,ProfileWSC,AutoFDO}.
PonyTail's control-flow tracing is not suitable for ``always on'' recording of production systems, due to its run-time overhead and data volume~(\cref{chap:eval:online}).
However, we believe the overhead is acceptable for intermittent or manual data collection over short time periods.
Disruption due to data collection can be further reduced by recording on many machines for very short time periods.

PonyTail's tracing analysis takes about 4.5--6 minutes per single-core second of service execution~(\cref{chap:eval:offline}).
Analysis time is mitigated by the fact that a few seconds of execution suffice for capturing tail events at microsecond-to-millisecond scale service times.
It is also possible to implement a PonyTail ``server'' that analyzes data once and answers queries about it.

Our PonyTail prototype targets C/C++ binaries and does not support services with self-modifying or just-in-time (JIT) compiled code (e.g., Java services).
Supporting JITed code is a matter of engineering effort, as \texttt{perf} can support JIT symbols with cooperation from the JIT compiler~\cite{PerfJIT}.

\hOne{Case studies}\label{chap:cases}

To demonstrate the merit of PonyTail's methodology and framework, this section presents several case studies in which we walk through applying PonyTail to analyze single-worker tail service time in five latency-critical services from the TailBench~\cite{tailbench-paper} benchmark suite~(\cref{sec:benchmark}).
Unless noted otherwise, each described analysis is based on a single PonyTail tracing data collection pass from the service's execution.

PonyTail's analysis reveals cases of \StCo and \uarch~(\cref{sec:method}) as reasons for service time outliers in some services and \DyCo in others.
Based on PonyTail's analysis, we develop tail service time optimizations~(\crefrange{subsec:silo}{subsec:moses}) and broad service time optimizations~(\crefrange{subsec:xapian}{subsec:sphinx}), and obtain other actionable insights~(\cref{subsec:shore}).
Finally, we reflect on the aspects of PonyTail that facilitated our analysis compared to standard hotspot-based profiling~(\cref{sec:comparison}).

\hTwo{Experimental setup}\label{sec:setup}

\para{Platform}
We use a server with an Intel Xeon Gold 6442Y (Sapphire Rapids) processor with 24 2.6\,GHz cores, 256\,GB DDR5-4800 memory, and a datacenter-class NVMe Samsung PM9A3~\cite{pm9a3} solid-state drive.
The server runs Ubuntu 20.04.3 (Linux kernel 6.4).
Code is compiled with GCC 9.4.0.

To control for external performance fluctuations, we~(1)~disable dynamic CPU frequency scaling, in both hardware (Turbo Boost) and the OS; (2)~minimize hardware resource sharing by disabling hyper-threading, pinning the service to a core, and using an otherwise idle system.

\para{Execution harness}
We use the common harness TailBench provides for statistically sound load testing of each service.
The harness accepts parameters controlling the workload: (1)~client requests per second (RPS); (2)~number of warmup requests, to ensure only steady-state execution is measured; and~(3)~number of requests to measure (measurement period).
PonyTail records only the measurement period.
Except for \sphinx, where we increase the measurement period to 20\,requests, we use TailBench's default parameters when analyzing the services~(\cref{tab:tbapps}).
We use low loads for the analysis but evaluate optimization impact across all loads.
Reported service/response times are the average of 3 runs.

The harness supports a range of scenarios, from a networked configuration to a single-node integrated configuration, in which the client and service are combined in a single process.
The client and server measure response and service time, respectively.
For simplicity, we use the integrated configuration, which was shown to accurately capture service time and latency characteristics~\cite{tailbench-paper}.
We use one server (worker) thread.
The server's request queue is managed with an FCFS policy.

Integrating PonyTail into TailBench requires changing fewer than a dozen LOC, to invoke PonyTail when the harness invokes the service and when the service responds.
One service (\shore) uses an internal work queue in addition to the system's request queue, so we invoke PonyTail from the service's worker thread, which does the actual work.

\para{Harness optimization}
We observe that the constructor of \texttt{Request} and \texttt{Response} objects in the harness redundantly zeroes out the objects which are later overwritten before being used.
This results in non-negligible overhead, particularly on services with low service time scales.
We resolve the problem by removing the superfluous zeroing, which, e.g., improved \silo's and \masstree's mean service time by $\approx 2\times$.

 \begin{table}
    \footnotesize
    \centering
    \begin{tabular}{p{3em}|p{8em}p{3em}||p{4em}p{3.6em}}
        \toprule
        & \multicolumn{2}{c}{\textbf{Description}} & \multicolumn{2}{c}{\textbf{Analysis parameters}} \\
        \textbf{Service} & \textbf{Domain} & \textbf{Service rate (RPS)} & \textbf{\#\,Warmup requests} & \textbf{\#\,Profiled requests} \\
        \midrule
        \silo & In-memory database &  66.7\,K & 20,000 & 20,000 \\
        \hline
        \moses & Statistical machine translation & 654 & 500 & 500 \\
        \hline
        \xapian & Online search & 1388 & 1000 & 3000 \\
        \hline
        \sphinx & Speech recognition &  2 & 10 & 20 \\
        \hline
        \shore & On-disk database & 2860 & 10,000 & 2000 \\
        \bottomrule
    \end{tabular}
    \caption{\textbf{Analyzed services.} Description, single-core service rate~($\mu$), and analysis evaluation parameters.}
    \label{tab:tbapps}
\end{table}

\hTwo{TailBench services}\label{sec:benchmark}

TailBench~\cite{tailbench-paper} provides a collection of eight latency-critical services.
These services span a wide range of application domains and service times, from microseconds through milliseconds to seconds.
Seven out of the eight services are C++ programs, of which we analyze the five services~(\cref{tab:tbapps}) that exhibit dispersive service times.
We provide more details on each service and its workload in the later sections describing their analyses.

\Cref{fig:xapian-time} shows the service time cumulative distribution function (CDF) of the services, which visualizes their high dispersion.
Except for one, all CDFs are similar in shape to those reported by TailBench~\cite{tailbench-paper}, although the magnitude of service times can differ due to our faster
hardware and hareness optimizations.
The only exception is \moses, whose significant tail service times are absent from the TailBench results.
We explain this discrepancy in~\cref{subsec:moses}.

\hTwo{Silo}\label{subsec:silo}

\Silo~\cite{silo-paper} is a fast in-memory transactional database system.
Transactional databases are widely used in online transaction processing (OLTP) systems.
TailBench drives \silo using the TPC-C workload~\cite{tpcc}, a standard e-commerce model.

\para{Characterization}
\Cref{fig:silo-char} shows PonyTail's characterization of \silo requests, indicating that service time variability is associated with both \StCo and \uarch.
The DBBL heatmap visualizing the code footprints~(\cref{fig:silo-char-hmap}) exhibits significant differences in executed code (locations and frequency) between different percentiles.
The TMA metrics~(\cref{fig:silo-char-tma}) show the same partition into groups, with each group having a distinct TMA breakdown.

DBBLs of each group correspond to handling of different request types.
For instance, STOCK\_LEVEL and DELIVERY transactions have DBBLs that are prevalent in 90--95th percentile and $>95$th percentile requests, respectively;
overall, despite each of these types being only 4\% of all transactions, they are the majority of the $>90$th percentile requests.
This is consistent with prior work that modeled these two transactions as the slowest in the TPC-C mix~\cite{Persephone}.
However, PonyTail's analysis reveals this characterization naturally, in a workload-agnostic way.
This can lead to finding ``implicit'' request types, i.e., properties of a request that predict its service time~(see~\cref{subsec:xapian,subsec:sphinx}).

Importantly, not \emph{all} of \silo's service time variability is due to different request types.
\Silo exhibits a long tail of requests whose service time exceeds the 99th percentile~(\cref{fig:silo-time}b).
\Cref{fig:silo-char-hmap} zooms in on the slowest 10 requests to show the code footprint of these long tail requests, which turns out to be part of \silo's garbage collection (GC) mechanism.

\begin{figure}
        \centering
        \includegraphics[width=0.5\columnwidth]{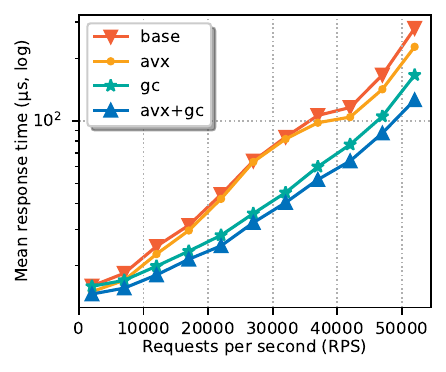}%
        \includegraphics[width=0.5\columnwidth]{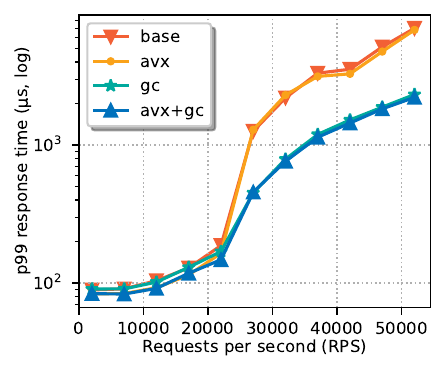}\\
        \begin{minipage}[t]{0.5\columnwidth}
        \centering
        \subcaption{Mean.} \label{fig:silo-speedup-mean}
        \end{minipage}%
        \begin{minipage}[t]{0.5\columnwidth}
        \centering
        \subcaption{99th percentile.} \label{fig:silo-speedup-p99}
        \end{minipage}%
        \caption{\textbf{Response time speedups (log scale) of optimizations on \silo}: \texttt{gc} and \texttt{avx} refer to~\cref{sec:silo-ctrl} and~\cref{sec:silo-uarch}, respectively.}
    \label{fig:silo-opt-times}
\end{figure}

In the following, we determine some causes for \silo's service time variability by analyzing the code footprint of the long tail requests~(\cref{sec:silo-ctrl}) and the microarchitectural footprint of STOCK\_LEVEL requests~(\cref{sec:silo-uarch}), where a significant fraction ($\approx 35$\%) of pipeline slots is wasted due to branch misprediction.
\Cref{fig:silo-opt-times} shows the impact of addressing these root causes on response time.
For instance, at \result{32\%} load (22\,KRPS) the mean, 95th, and 99th percentile response times improve by up to \result{43\%}, \result{12\%}, and \result{21\%}.

\hThree{Code footprint exploration} \label{sec:silo-ctrl}

\Silo's DBBL heatmap~(\cref{fig:silo-char-hmap}) shows that the slowest 10 requests execute DBBL $\tau$=\texttt{silo+0x106561} which is part of the \texttt{clean\_up\_to\_incl\-uding} method.
We use PonyTail's DBBL-Info report, restricted to requests that execute $\tau$, to identify all DBBLs that have $>10\%$ hotness in such requests.
The result~(\cref{table:silo-gc-bbl-info}) shows that in these requests, 81\% of the time is spent in \texttt{clean\_up\_to\_incl\-uding} DBBLs.
Moreover, request service times are highly correlated with both execution count (0.79) and time (0.93) of $\tau$, which is also the hottest DBBL.

\begin{table}[t]
\centering
\footnotesize
\begin{tabular}{l l | l l l r}
\toprule
\textbf{DSO} & \textbf{Offset} & \textbf{Depth} & \textbf{\# Execs} & \textbf{Hot} & \textbf{Time ($\mu$s)} \\
\midrule
\silo & 0x106561 & 3 & 18046.30 (0.79) & 0.45 & 43.01 (0.93) \\
\multicolumn{6}{l}{\texttt{transaction\_proto2\_static::clean\_up\_to\_including(...)+0x261}} \\
\silo & 0x106522 & 3 & 17250.30 (0.79) & 0.25 & 24.49 (0.72) \\
\silo & 0x106658 & 3 & 796.00 (0.75) & 0.11 & 10.37 (0.88) \\
\bottomrule
\end{tabular}
\caption{\textbf{PonyTail DBBL-Info report} (for \silo requests where \texttt{clean\_up\_to\_including()} is invoked). Correlations with service time are shown in parentheses. The actual report includes symbol names for every line.}
\label{table:silo-gc-bbl-info}
\end{table}

The method \texttt{clean\_up\_to\_including} is called by \silo's garbage collection (GC) mechanism, which uses an epoch-based memory reclamation scheme in the style of Read-Copy Update (RCU)~\cite{McKenneyS1998,fraser-phd}.
This GC is invoked by \silo worker threads periodically to reclaim unreferenced objects that have accumulated since the last GC.
Increasing the GC's frequency will cause more requests to be slowed down by GC, but each GC execution will be faster as it will find fewer objects to reclaim.
We therefore hypothesize that this will improve tail service time.
We increase GC frequency by $5\times$.
This change reduces the maximum and 99th percentile service times at \result{32\%} load by \result{$4.3\times$} and \result{1\%}, respectively.

\hThree{Microarchitectural footprint exploration} \label{sec:silo-uarch}

We pinpoint the sources of branch misprediction in STOCK\_LEVEL transactions by running PonyTail's data collection using sampling based on the \texttt{TOPDOWN.BR\_MISPREDICT\_SLOTS} performance counter and filtering the results by the existence of STOCK\-\_LEVEL DBBLs in the sampled stack.
We find that most branch mispredictions occur during searches in \silo indexes, which are based on Masstree~\cite{masstree-paper}.
Specifically, 32\% of samples occur in the code that determines the next node on the search path.
This code performs a binary search in the current node's key array to locate the first key greater than or equal to the target key.

We first attempt to replace the binary search branches with conditional move (\texttt{CMOV}) instructions, which set the left/right boundaries of the search based on the current key comparison.
However, this change does not improve service time, and repeating PonyTail's characterization reveals that the bottleneck for this request group has shifted to the CPU backend.
We therefore perform a different optimization.
We use the fact that the key array contains 15 8-byte values to execute all comparisons with a constant number of AVX-512~\cite[\S~15]{sdm-manual} vector instructions.
We load the array into two AVX-512 registers, obtain a 15-bit mask indicating the comparison result of each index, and return the index of the first set bit (using the \texttt{TZCNT} instruction).

As a result, the 95th percentile service time at \result{32\%} load improves by \result{12\%}.

\hTwo{Moses}\label{subsec:moses}

The \moses service performs statistical machine translation~\cite{moses-paper, moses-website}, modeling the services underpinning online translation and speech-based applications~\cite{tailbench-paper}.
It is queried on randomly chosen dialogue snippets from the \url{opensubtitles.org} English-Spanish corpus~\cite{moses-queries}.

In the following, we identify the root cause of \moses' long tail (beyond the 99th percentile) service times~(\cref{fig:moses-time}c).
Addressing this cause (by modifying the benchmark configuration) reduces the maximum service and response times by $53\times$ and $22\times$, respectively.

\begingroup
\setlength{\columnsep}{9pt}

\begin{wrapfigure}[21]{r}{0.5\columnwidth}
        \centering
        \vspace{-20pt}
        \includegraphics[width=0.5\columnwidth]{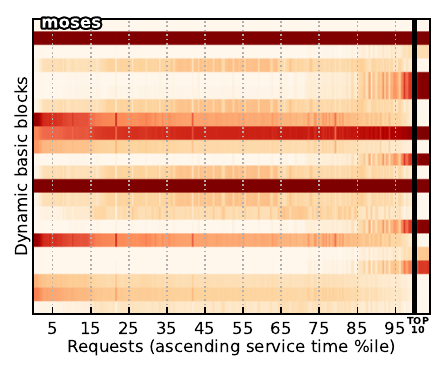}\\
        \vspace{-10pt}
        \caption{\textbf{PonyTail analysis of \moses requests:} $\left<20\%,1,10\right>$-DBBL heatmap.} \label{fig:moses-char-hmap}
    \label{fig:moses-char}
\end{wrapfigure}
\para{Characterization}
\Cref{fig:moses-char} shows PonyTail's characterization of \moses requests.
The code footprints are similar across almost all requests, except for the slowest requests that exhibit a distinct pattern.
The microarchitectural footprints do not exhibit interesting trends and are omitted.

\endgroup

Code inspection of the DBBLs associated with the slowest requests shows that they are related to failed lookups in a ``translation cache'' structure,
which stores source and target phrases and is persistent across requests.

Suspecting these failures are related to an insufficiently warmed-up cache, we re-analyze \moses with 5000 warmup requests instead of 500.
Consequently, the offending DBBLs disappear from PonyTail's code footprint report.
The maximum service and response times are reduced by $53\times$ and $22\times$.

This observation highlights the importance of the translation cache in \moses.
It can also explain the difference between our service time CDF for \moses and the TailBench CDF, if the default value in the TailBench code differs from the value used to generate the CDF in the TailBench paper.

\hTwo{Xapian}\label{subsec:xapian}

TailBench's \xapian service represents a node in an online search engine, using the Xapian open-source search engine library~\cite{xapian-website}.
An online search engine spreads petabytes of index data across thousands of such nodes, each of which searches its portion of the index upon a user query.
\Xapian indexes a dump of the English Wikipedia and is queried on English words and names chosen randomly from a Zipfian distribution, which mirrors the patterns of online search queries.
For each query term specified in a request, \xapian responds with a set of results from its index.

\begin{figure}
        \centering
        \includegraphics[width=0.5\columnwidth]{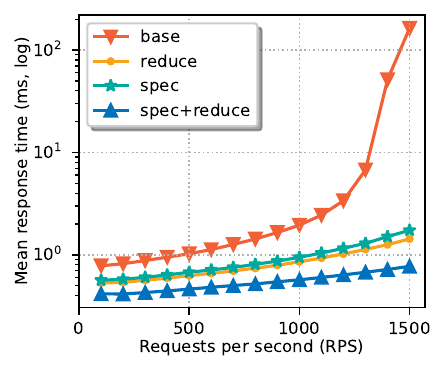}%
        \includegraphics[width=0.5\columnwidth]{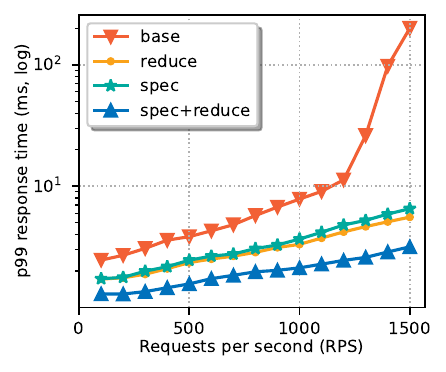}\\
        \begin{minipage}[t]{0.5\columnwidth}
        \centering
        \subcaption{Mean.} \label{fig:xapian-speedup-mean}
        \end{minipage}%
        \begin{minipage}[t]{0.5\columnwidth}
        \centering
        \subcaption{99th percentile.} \label{fig:xapian-speedup-p99}
        \end{minipage}%
        \caption{\textbf{Response time speedups (log scale) of optimizations on \xapian}: \texttt{reduce} and \texttt{spec} refer to~\cref{sec:xapian-ctrl} and~\cref{sec:xapian-uarch}, respectively.}
    \label{fig:xapian-opt-times}
\end{figure}

We identify control-flow~(\cref{sec:xapian-ctrl}) reasons for service time variability in \xapian and determine their root causes, and identify a subtle CPU bottleneck~(\cref{sec:xapian-uarch}).
Addressing these issues improves \xapian's service and response times across the board~(\cref{fig:xapian-opt-times}  and thus the maximal sustainable load as well.
For instance, at \result{87\%} load (1200\,RPS) the mean and 99th percentile response times improve by \result{$5.3\times$} and \result{$4.6\times$}, respectively.

\begin{figure}
        \centering
        \includegraphics[width=0.5\columnwidth]{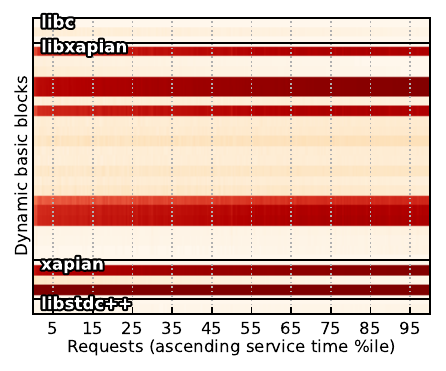}%
        \includegraphics[width=0.5\columnwidth]{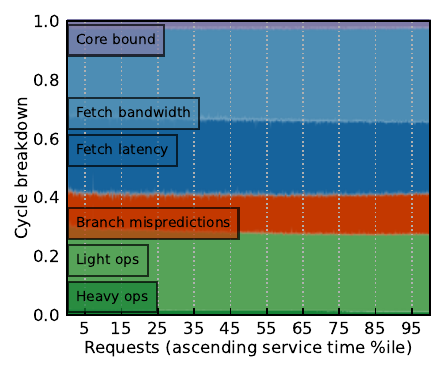}\\
        \begin{minipage}[t]{0.5\columnwidth}
        \centering
        \subcaption{$\left<10\%,1\right>$-DBBL heatmap.} \label{fig:xapian-char-hmap}
        \end{minipage}%
        \begin{minipage}[t]{0.5\columnwidth}
        \centering
        \subcaption{TMA metrics.} \label{fig:xapian-char-tma}
        \end{minipage}%
        \caption{\textbf{PonyTail analysis of \xapian requests.}}
        \label{fig:xapian-char}
\end{figure}

\begingroup
\setlength{\columnsep}{9pt}

\para{Characterization}
\Cref{fig:xapian-char} shows PonyTail's characterization of \xapian requests.
The DBBL heatmap~(\cref{fig:xapian-char-hmap}) exhibits no significant difference in executed code (locations or frequency) between different percentiles (i.e., no \StCo).
Likewise, TMA metrics~(\cref{fig:xapian-char-tma}) show no unique characteristics of high service times (i.e., little \uarch),
\begin{wrapfigure}[14]{r}{0.5\columnwidth}
    \begin{center}
    \vspace{-9pt}
    \includegraphics[width=0.5\columnwidth]{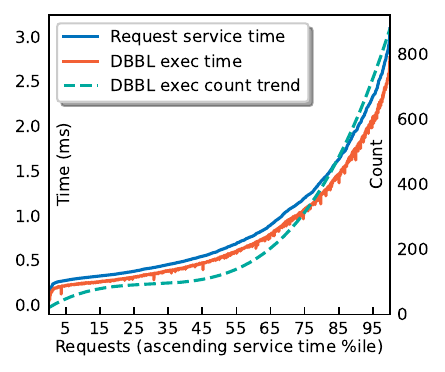}
    \end{center}
    \vspace{-10pt}
    \caption{\textbf{PonyTail DBBL-Follow report of \xapian key DBBLs:} Correlation between service time and execution count of DBBL.} \label{fig:xapian-bbl}  
    \label{fig:xapian-bbl}
\end{wrapfigure}
except for a slight increase in the fraction of pipeline slots unutilized due to fetch bandwidth and branch misprediction.
We therefore classify all requests into one group and conclude that \xapian's tail latency is due to \DyCo, i.e., some request DBBLs execute more in tail requests.
We use PonyTail's DBBL-Info report to identify the \emph{key DBBL}---the highest DBBL in the service call stack which~(1)~executes multiple times during request handling and~(2)~whose number of executions and time spent in its call stack are correlated with request service time.
\Cref{fig:xapian-bbl} depicts the PonyTail DBBL-Follow report of the key DBBL, showing the correlation of its execution time and count with service time.

\endgroup

\hThree{Code footprint exploration} \label{sec:xapian-ctrl}
We examine the source code associated with the key DBBL to understand its execution context.
The DBBL executes once per document found for the searched term, as part of a loop that iterates over a list of objects, each of which represents a document indexed by the searched term.
For each such document, the loop invokes a \texttt{get\_weight()} method, part of which looks up the document in \xapian's database backend.
The backend maintains fixed-size \emph{chunks}, each of which describes a number of documents, consecutive by their IDs.
To retrieve information for a document, the backend first locates the relevant chunk and then traverses it to find the desired document's data.
Locating the relevant chunk consists of checking if the last inspected chunk is a match, and moving to the correct chunk otherwise.

PonyTail's DBBL-Info report shows that, on average, moving to a different chunk is done for $\approx 166$ documents out of the $\approx 230$ scanned and takes 11\% of service time, while locating the correct position inside the chunk takes 74\% of service time.

We hypothesize that it will be profitable to reduce the backend's chunk size, as~(1)~searching inside a chunk is more expensive than locating a chunk, and~(2)~locating chunks is already done frequently.
To test this hypothesis, we evaluate \xapian on the same database built with 1\,KB chunks instead of the default 2\,KB.
This change reduces mean and tail service times by \result{$\approx 30$\%}, improving mean and tail response times at \result{87\%} load by \result{$\approx 3\times$}.

\hThree{Microarchitectural footprint exploration} \label{sec:xapian-uarch}

\Cref{fig:xapian-char-tma} shows that the majority of pipeline slot stalls in \xapian request handling are caused by CPU frontend issues: branch mispredictions (26\%) and fetch bandwidth underutilization (31\%), which refers to the frontend delivering fewer $\mu$-ops to the backend than its capacity.
Intel CPUs provide numerous performance counters for tracking events responsible for fetch bandwidth underutilization.
To investigate, we run PonyTail data collection again, this time collecting the request-level deltas of these performance counters.

The output points to an issue with utilization of the $\mu$-op cache, also called the decoded stream buffer (DSB).
The DSB is a set-associative cache that caches the $\mu$-ops of decoded instructions, thereby allowing the frontend to skip decoding of DSB-cached instructions~\cite[\S~E2.2.2]{opt-manual}.
Compared to the standard instruction decoding path, $\mu$-op delivery from the DSB improves throughput, saves power, and reduces latency.

PonyTail's report indicates that, on average per request, $< 3\%$ of retired instructions experienced a DSB miss and that the DSB delivers $\mu$-ops during most cycles---but for $\approx 47\%$ of all cycles, the DSB delivers a suboptimal number of $\mu$-ops.
To observe this, we use the counters \texttt{IDQ.DSB\_CYCLES\_ANY} and \texttt{IDQ.DSB\_CYCLES\_OK}, which count the cycles in which the DSB delivers any $\mu$-ops and the optimal number of $\mu$-ops, respectively.
By running PonyTail's data collection using sampling based on these performance counters, we pinpoint the problem to the chunk searching code of \xapian's backend.

As the relevant code scans the chunk looking for a specific document ID, it frequently invokes an \texttt{unpack\_uint} method~(\cref{fig:chert-src}),
which decodes an unsigned integer from a string form in the chunk buffer.
Out of the 8 most commonly sampled instructions, 6 are branch instructions related to this code.
Now, according to Intel's documentation~\cite{opt-manual}, each DSB entry caches (up to 6) $\mu$-ops of statically contiguous instructions within some 32-byte aligned region; at most two branches are allowed per entry; and a non-conditional branch terminates the cached sequence.
These restrictions mean that DSB entries of branch-heavy contiguous code sequences will be underutilized.

Using PonyTail's Loop-Follow report on the \texttt{unpack\_uint} loop, we observe that at most 4 iterations of the loop are executed.
This observation gives rise to the following optimization: unroll the first 4 iterations of the loop~(\cref{fig:xapian-chunk-no-loop}), so that~(1)~basic blocks likely to execute sequentially are laid out contiguously in memory and~(2)~we avoid the \texttt{while} branches of the original loop.
This change improves the fraction of DSB-optimal delivery cycles by $\approx 4\times$.
Consequently, mean and tail service times are reduced by \result{$\approx 27$\%}, which improves mean and tail response times at \result{87\%} load by \result{$2.9\times$} and \result{$2.3\times$}.

\begin{figure}
\begin{minipage}[t]{.5\columnwidth}
\begin{lstlisting}[style=C,numbers=none]
...
 // Check the length of the encoded integer first.
 do {
  if (rare(ptr == end)) {
   // Out of data.
   *p = NULL;
   return false;
  }
 } while (*ptr++ >= 128);

...
@(* *)@
@(* *)@
\end{lstlisting}
\end{minipage}%
\begin{minipage}[t]{.5\columnwidth}
\begin{lstlisting}[style=C,numbers=none]
if (ptr == end) {
 *p = NULL;
 return false;
} else if (!(*ptr++>=128)) {
 *p = ptr;
} else if (ptr == end) {
 *p = NULL;
 return false;
} else if (!(*ptr++>=128)) {
 *p = ptr;
}
// (repeat ptr check again twice)
// Original loop (fallback)
...
\end{lstlisting}
\end{minipage}\\
\begin{minipage}[t]{.5\columnwidth}
\subcaption{Original \texttt{unpack\_uint} code.}
\label{fig:chert-src}
\end{minipage}%
\begin{minipage}[t]{.5\columnwidth}
\subcaption{Optimized code.}
\label{fig:xapian-chunk-no-loop}
\end{minipage}%
\caption{\textbf{Optimization for better \xapian $\mu$-op cache throughput.}}
\end{figure}

\hTwo{Sphinx}\label{subsec:sphinx}

The \sphinx service is an accurate speech recognition system~\cite{sphinx-paper-1, sphinx-paper-2, sphinx-website}, which is a key component in speech-based applications.
TailBench queries \sphinx on random audio utterances from the CMU AN4 database~\cite{CMUAN4} and the service responds with the hypothesis text.
Request handling is compute-intensive, with a mean service time of $\approx 0.5$ seconds.

We identify the root cause of \sphinx's service time variability, as well as a general optimization opportunity to alleviate a microarchitectural bottleneck in all requests.
\Cref{tab:sphinx-speedup} shows the optimization's impact at \result{50\%} load, with across the board improvements of service and response times.

\begin{table}
    \footnotesize
    \centering
    \begin{tabular}{c|cccc}
        \toprule
        \textbf{Time} & \textbf{Latency} & \textbf{Base} & \textbf{Optimized} \\
        \midrule
        Service & Mean & \result{525\,ms} & \result{$-10.7$\%} \\
          & 95th percentile & \result{948\,ms} & \result{$-10$\%} \\
          & 99th percentile & \result{1097\,ms} & \result{$-10.7$\%} \\
        Response & Mean & \result{747\,ms} & \result{$-15.2$\%} \\
          & 95th percentile & \result{1325\,ms} & \result{$-16$\%} \\
          & 99th percentile & \result{1700\,ms} & \result{$-14.5$\%} \\
        \bottomrule
    \end{tabular}
    \caption{\textbf{Impact of PonyTail-assisted optimization on \sphinx (1\,RPS$=$50\% load).}}
    \label{tab:sphinx-speedup}
\end{table}

\para{Characterization}
Similarly to \xapian~(\cref{subsec:xapian}), \sphinx's code footprints and TMA metrics are similar across service times (so we omit them).
We therefore classify all requests into a single group, again concluding that \sphinx's tail latency is due to \DyCo.
As with \xapian, we use PonyTail's DBBL-Info report to identify the key DBBL that is correlated with tail service time.
We find that correlation between the key DBBL's execution time and service time exists for all requests, but correlation between its execution count and service time is most noticeable for tail requests.

Code inspection shows that the execution count of the key DBBL is tied to the number of audio samples in the input utterance.
Identifying such a request-level predictor of service time (an ``implicit type'') is important, as it can be leveraged by request schedulers to improve overall tail latency~\cite{Persephone}.

\para{Optimization}
Using the same process as in \silo's microarchitectural footprint exploration~(\cref{sec:silo-uarch}), we pinpoint the sources of branch misprediction in \sphinx.
The culprit is a loop evaluating a Gaussian Mixture Model
(GMM)~\cite{GMM} which computes the likelihood that an observation vector belongs to one of several clusters represented as Gaussian distributions.
The likelihood is computed incrementally over the vector dimensions, with the loop terminated early if the likelihood crosses some threshold.
The resulting loop termination condition requires multiple branches to evaluate.
Each loop iteration computes over 4 dimensions, in an apparent attempt to encourage compiler vectorization.

PonyTail's Loop-Follow report shows that this loop iterates over at most 13 dimensions at run-time.
We thus change the unrolling factor to 13.

This change significantly reduces the number of branches (and mispredictions) during the loop and improves service time by \result{$\approx 10$\%}, despite performing more wasted work (as the early termination condition is now checked less frequently).

\hTwo{Shore}\label{subsec:shore}

\Shore~\cite{shore-paper} is an on-disk transactional database, which differs from \silo in how it stores data.
To avoid disk I/O bottlenecks, we use a solid-state drive for \shore's database and logs.
Similarly to \silo, it is driven by TPC-C.

\para{Characterization}
\Shore's code footprints reveal several request groups, which all exhibit relatively similar microarchitectural footprints (we omit the figures).

Similarly to \silo, \shore's request groups are caused by different transaction types.
However, \shore's service time trends differ from \silo's.
While most requests above the 95th percentile involve STOCK\_LEVEL transactions, we observe NEW\_ORDER transactions all across the 50th--96th percentiles and even in the slowest 10 requests.

This phenomenon underscores the importance of understanding service performance, as it can be a determining factor even for broadly used and well-understood workloads.

\hTwo{Reflections} \label{sec:comparison}

Given that PonyTail records the same data as Linux's \texttt{perf}, it is natural to ask whether our optimization effort could be carried out using \texttt{perf}.
Here, we pinpoint PonyTail's unique aspects that enable the analyses.
Improving service time in \silo and \moses required identifying code and microarchitectural behaviors that occur only in outlier requests.
These do not show up (or are ranked low) on a standard \texttt{perf} report.
In contrast, in \xapian and \sphinx we improve service time across the board by addressing a global hotspot.
However, successful optimization requires more than identifying the bottleneck.
In \xapian, PonyTail's DBBL-Info report is necessary for the control-flow optimization~(\cref{sec:xapian-ctrl}).
It shows that locating chunks is already done for 72\% of the documents yet only takes 11\% of request time, which leads to the conclusion that trading off more chunk-locating for fewer chunk-searches will improve service time.
For the microarchitectural optimization~(\cref{sec:xapian-uarch}), PonyTail's Loop-Follow is crucial.
It enables determining that each execution of the hot loop with poor DSB utilization runs for at most 4 iterations, which leads to the unrolling optimization.
Loop-Follow is similarly crucial for the \sphinx optimization~(\cref{subsec:sphinx}).

We also consider Intel's VTune Profiler~\cite{VTune}.
It has recently introduced a preview Anomaly Detection analysis feature, which targets analyzing performance outliers---it allows the user to group requests by latency and observe their performance data for these groups.
However, VTune's analysis is orders of magnitude slower than PonyTail's~(\cref{chap:eval:offline}), taking minutes per millisecond.
In addition, data collection is limited to 5000 requests of at most 1 second each.
This limitation restricts the scope of analysis, especially for services with high request volume and variety, such as \silo and \shore.
PonyTail has no similar limitation.
VTune only allows comparing performance metrics of two specific requests at a time, which hinders broad characterization such as PonyTail's code and microarchitectural footprint visualization.
Finally, VTune cannot perform analysis at request group level, which is a PonyTail functionality heavily used in our analyses.

\hOne{Evaluation}\label{chap:eval}

We evaluate PonyTail's online overhead, during program execution~(\cref{chap:eval:online}), and offline (analysis) efficiency~(\cref{chap:eval:offline}).
We use the setup described in~\cref{sec:setup}.
We report averages of 3 runs; if variance is not negligible, we include error bars showing the standard deviation.

\hTwo{Run-time overhead} \label{chap:eval:online}

\begin{figure}[t]
\begin{minipage}[t]{\columnwidth}
    \begin{minipage}[t]{0.5\columnwidth}
    \centering
    \includegraphics[width=\textwidth]{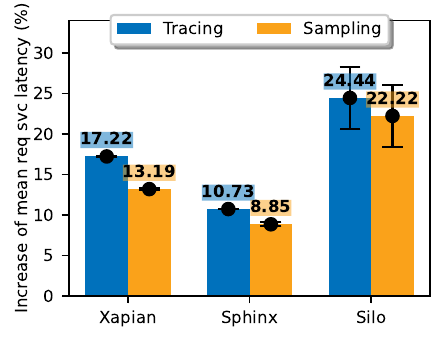}
    \end{minipage}%
    \begin{minipage}[t]{0.5\columnwidth}
    \centering
    \includegraphics[width=\textwidth]{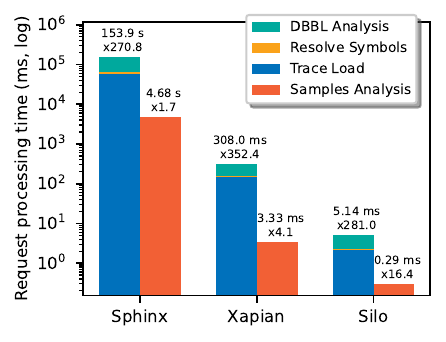}
    \end{minipage}\\
    \begin{minipage}[t]{0.48\columnwidth}
    \centering
    \captionof{figure}{Data collection overhead.}
    \label{fig:runtime-overhead}
    \end{minipage}\,\,\,\,
    \begin{minipage}[t]{0.48\columnwidth}
    \centering
    \captionof{figure}{Per-request analysis time (absolute and relative to mean service time).}
    \label{fig:analysis-time}
    \end{minipage}
\end{minipage}
\end{figure}

\para{Impact on service time}
\Cref{fig:runtime-overhead} shows the relative increase in mean service time of \xapian, \sphinx, and \silo for each PonyTail data collection mode.
For sampling mode, we sample every 100\,K cycles.
PonyTail's overhead varies across both services and modes, with tracing and sampling imposing a geometric average slowdown of 16.5\% and 13.7\%, respectively.
Tracing overhead mainly consists of PT recording overhead (on all requests) and PT buffer-full interrupts (on interrupted requests).
Similarly, sampling overhead consists of PEBS overhead and interrupts, whose frequency depends on the sampling parameters and program execution.

For both modes, PonyTail's relative overhead decreases with service time, but not in direct proportion.
PonyTail's relative overhead is only $2.4\times$ higher in \silo than in \sphinx, despite \silo requests being orders of magnitude faster than \sphinx requests.
Because PonyTail uses \texttt{perf}'s data collection mechanism, its overheads are comparable to those of standard profiling with \texttt{perf}'s analogous modes.

\para{Impact on service time distribution}
We verify that service time distributions remain similar in shape (only shifted due to the overhead) when PonyTail data collection is running.
This ensures that PonyTail does not misclassify non-tail requests as tail requests or vice versa.

\hTwo{Analysis efficiency} \label{chap:eval:offline} \label{subsec:analysis}

\para{Time}
\Cref{fig:analysis-time} shows PonyTail's per-request analysis time of both trace and sample data for \xapian, \sphinx, and \silo.
Tracing analysis time scales linearly with service time, reflecting a rate of 4.5--6 minutes per second of service execution.
Total analysis time depends on the amount of data collected per request---so trace analysis is $60\times$ slower on average than sampling analysis---and on the number of requests recorded.
For our parameters~(\cref{tab:tbapps}), total trace analysis times are 12.6 minutes, 45 minutes, and 15 seconds, respectively.

\begingroup
\setlength{\columnsep}{10pt}

\begin{wraptable}[7]{r}{0.22\textwidth}
    \centering
    \footnotesize
    \vspace{-10pt}
    \begin{tabular}{l|ccc}
    \toprule
    \textbf{Service} & \textbf{Tracing} & \textbf{Sampling} \\
    \midrule
    Xapian & 260\,GB & 2\,GB \\
    Sphinx & 469\,GB & 6\,GB \\
    Silo   & 18\,GB & 2\,GB \\
    \bottomrule
    \end{tabular}
    \caption{PonyTail analysis memory consumption.}
    \label{tab:memory_consumption}
\end{wraptable}
\para{Memory usage}
\Cref{tab:memory_consumption} reports PonyTail's analysis peak virtual memory usage when analyzing trace or sampling data.
Processing of trace data utilizes up to hundreds of GBs of virtual memory, while memory usage during sample processing is on a smaller scale of a few GBs.

\endgroup

\para{Takeaways}
PonyTail's trace analysis is memory intensive and, depending on service times, can be time consuming as well.
Fortunately, however, the more prevalent microsecond to millisecond service times result in negligible to moderate analysis time.
Moreover, our experience is that analysis time is not the bottleneck in the service time optimization work flow---obtaining insights and developing optimizations based on them form the lion's share of the work.

\hOne{Related work}\label{sec:literature}

We first place PonyTail's conceptual contribution in context and then discuss its technical aspects.

\para{Profilers}
General-purpose profilers report performance prob\-lems---such as execution hotspots~\cite{PerfTool,VTune,OptiWISE}, high data access latency~\cite{DissectingMemAcc,CachePerf} and bandwidth~\cite{PerfMemPlus}, or useless work~\cite{JXPerf,LDoctor}---but do not identify their root causes.
Specialized profilers locate root causes of specific issues, e.g., slow execution~\cite{VACostProfiling} or lock contention~\cite{SyncPerf,FreeLunch,wPerf}.
PonyTail sits midway between these approaches: it does not identify root causes, but has features to assist root-cause analysis.

\para{Automated profile-guided service optimization}
Several systems profile datacenter services in production to perform automated profile-guided optimization, whether online~\cite{OCOLOS} or offline~\cite{DMon,AutoFDO,APT-GET,BOLT,Propeller,Whisper,MaPHeA,CRISP}.
They do not target tail service time or performance outliers.
More importantly, PonyTail is complementary to this line of work.
The design of an automated optimization system starts with humans characterizing a class of performance issues and developing techniques to automatically fix them.
PonyTail targets this first step for the novel problem of specifically improving tail service time.

\para{Tail service time optimization}
Several works proposed techniques for improving tail service time in key-value stores~\cite{Rein,SILK+,SizeAwareSharding}.
PonyTail's goal is to assist such manual, service-specific efforts.
PonyTail is complementary to run-time systems techniques for speeding up slow requests, e.g., using parallelism~\cite{FewToMany,TPC} or voltage/frequency boosting~\cite{EETL,Adrenaline}.

\para{Execution time variance}
Prior work on analyzing execution time variance does not target PonyTail's more demanding problem of capturing any program- or CPU-level cause of tail service time.
For example, Hindsight~\cite{Hindsight} traces high-level program-defined events to troubleshoot tail latency in distributed systems, and is therefore not appropriate for understanding CPU or control-flow factors.
FVSampler~\cite{FVSampler} and Akiyama et al.~\cite{DiagFluct} synchronize profiling samples with function or request boundaries to profile their execution time variance, but they rely exclusively on performance counters and so cannot provide fine-grained control-flow data.
NSight~\cite{NSight} reconstructs network-message lifetimes from CPU and NIC hardware traces.
It targets network/OS latency, not tail service time, and cannot reveal intra-function factors.

\para{Cross-execution comparisons}
Comparison of multiple recorded executions is done by many systems for root-cause diagnosis or reproduction of software bugs~\cite{FailureSketching,LazyDiagnosis,ER,AURORA}.
Perspect~\cite{RelDbg} pinpoints the root cause of performance problems by comparing execution with and without the issue, but it
reasons exclusively about control-flow and so cannot reveal hardware-related issues.
Coz~\cite{Coz}
pinpoints promising code locations for optimization by running performance experiments during program execution.
However, while it supports average-latency as an optimization goal, it has no concept of tail optimization.

\hOne{Conclusion}\label{chap:summary}

This \work shows the opportunity to improve datacenter service tail latency by improving tail service time.
To unlock this potential, we present PonyTail, a performance analysis methodology and toolkit for analyzing service-level outlier behaviors associated with tail service time.
PonyTail helps performance engineers to identify control-flow and microarchitectural patterns responsible for service time outliers, as well as their root causes.
We demonstrate PonyTail's merit by optimizing several latency-critical services, improving their 99th percentile service time and latency by \result{10.7\%--46\%} and \result{21\%--$5\times$}, respectively.

\bibliographystyle{ACM-Reference-Format}
\bibliography{paper}

\appendix
\onecolumn

\section{DBBL extraction algorithm} \label{sec:appendix}

\begin{figure*}[h!]
\begin{python}
# enters, exits - containers which keep track of request-level counters for offsets
# ft_offsets - container of possible fall-through offsets
for (request, source, insn_len, target, br_type) in taken_branches:
    if br_type in {CALL, JMP, JCC}:
        enters[target][request]++
    if br_type in {RET, JMP, JCC}:
        exits[source + insn_len][request]++
    if br_type is JCC:
        ft_offsets.append(source + insn_len)

for ft_offset in ft_offsets:
    enters[ft_offset] # add as an enter offset, without increasing the counter

sort_descending(enters)
sort_descending(exits)
enter_idx = 0
# dbbls - container of dynamic basic blocks
# ft_count - container of request-level counters

for (exit_offset, exit_count) in exits:
    while (true):
        (enter_offset, enter_count) = enters[enter_idx]
        if (enter_offset < exit_offset):
            break;
        dbbls.append(enter_offset, enters[enter_idx - 1].offset, ft_count)
        enter_idx++
        ft_count -= enter_count

    if ft_offsets.contains(exit_offset):
        dbbl_count = (ft_count + exit_count)
    else:
        dbbl_count = exit_count

    dbbls.append(enter_offset, exit_offset, dbbl_count)
    enter_idx++
    ft_count -= enter_count
\end{python}
\caption{Algorithm for extracting DBBLs and their execution counts, at request-level, based on PT records of taken branches.}
\label{fig:bbl-alg}
\end{figure*}

\end{document}